\documentclass[aps,prb,showpacs]{revtex4}

\usepackage{graphicx}
\usepackage{amsfonts}

\newcommand{\nn}{\nonumber}

\newcommand{\tg}{\tilde{g}}
\newcommand{\la}{\langle}
\newcommand{\ra}{\rangle}
\newcommand{\bea}{\begin{eqnarray}}
\newcommand{\eea}{\end{eqnarray}}

\begin{document}

\title{Theory of Josephson Arrays in a Resonant Cavity}

\author{E. Almaas}
\email{Almaas.1@nd.edu}
\affiliation{Department of Physics, University of Notre Dame, Notre Dame, 
Indiana 46556}

\author{D. Stroud}
\email{stroud@mps.ohio-state.edu}
\affiliation{Department of Physics, The Ohio State University, Columbus,
Ohio 43210}

\date{\today}

\begin{abstract}
We review our previous work on the dynamics of one- and
two-dimensional arrays of underdamped Josephson junctions placed in a
single-mode resonant cavity.  Starting from a well-defined model
Hamiltonian, which includes the effects of driving current and
dissipative coupling to a heat bath, we write down the Heisenberg
equations of motion for the variables of the Josephson junction and
the cavity mode.  In the limit of many photons, these equations reduce
to coupled ordinary differential equations, which can be solved
numerically.  We present a review of some characteristic numerical
results, which show many features similar to experiment.  These
include self-induced resonant steps (SIRS's) at voltages $V =
n\hbar\Omega/(2e)$, where $\Omega$ is the cavity frequency, and $n$ is
generally an integer; a threshold number $N_c$ of active rows of
junctions above which the array is coherent; and a time-averaged
cavity energy which is quadratic in the number of active junctions,
when the array is above threshold.  When the array is biased on a
SIRS, then, for given junction parameters, the power radiated into the
array varies as the square of the number of active junctions,
consistent with expectations for coherent radiation.  For a given
step, a two-dimensional array radiates much more energy into the
cavity than does a one-dimensional array.  Finally, in two dimensions,
we find a strong polarization effect: if the cavity mode is polarized
perpendicular to the direction of current injection in a square array,
then it does not couple to the array and no power is radiated into the
cavity.  In the presence of an applied magnetic field, however, a mode
with this polarization would couple to an applied current.  We
speculate that this effect might thus produce SIRS's which would be
absent with no applied magnetic field.
\end{abstract}


\maketitle

\section{Introduction}

The properties of arrays of Josephson junctions have been of great
interest for nearly twenty years.\cite{review} These arrays are
excellent model systems in which to study phenomena such as phase
transitions and quantum coherence in two dimensions.  For example, if
only the Josephson coupling energy is considered, and if the
self-inductance and mutual inductance of the array plaquettes are
neglected, the Hamiltonian of a two-dimensional (2D) array of
Josephson junctions is formally identical to that of a 2D XY model
[see, e.\ g. Ref.\ \onlinecite{chaikin}].  In the presence of such
inductive effects, this XY description needs to be modified, and
several generalizations, which include these effects, have been
proposed.\cite{nakajima,majhofer,phillips,petraglia} Arrays sometimes
appear to mimic behavior seen in nominally homogeneous materials, e.\
g. high-$T_c$ superconductors, which often behave as if they are
composed of distinct superconducting regions linked together by
Josephson coupling.\cite{tinkham} Finally, the arrays are of
potentially practical interest: they may be useful, for example, as
sources of coherent microwave radiation if the individual junctions
can be caused to oscillate in phase in a stable manner.

Recently, our ability to achieve this kind of stable oscillation, and
coherent microwave radiation, was significantly advanced by a series
of experiments by Barbara and
collaborators.\cite{barbara,barbara02,asc2000,vasilic02,vasilic} These
workers placed two-dimensional underdamped Josephson arrays in a
resonant microwave cavity.  The presence of the cavity caused the
junctions to couple together with remarkable efficiency, resulting in
a highly efficient conversion of the injected d. c. power into
a. c. radiation.  Even more surprising, this efficiency is achieved in
{\em underdamped} arrays, which according to conventional wisdom,
should be especially difficult to synchronize, since each such
junction is both bistable and hysteretic.  Numerous workers have
attempted to explain these experiments.
\cite{filatrella,filatrella2,almaas,almaas02_2}

In this brief review, we summarize our own model to describe this kind
of phase locking.\cite{almaas02_2,almaas03}  In Section II, we present
the model Hamiltonian.  In Section III, we describe the equations of
motion resulting from this model, in the ``classical limit,'' as
defined further below. In Section IV, we give a brief description of
the relevant numerical results.  Finally, in Section V, we present a
concluding discussion.

\section{Model Hamiltonian}

\begin{figure}[pth]
\centerline{\includegraphics[height=6cm]{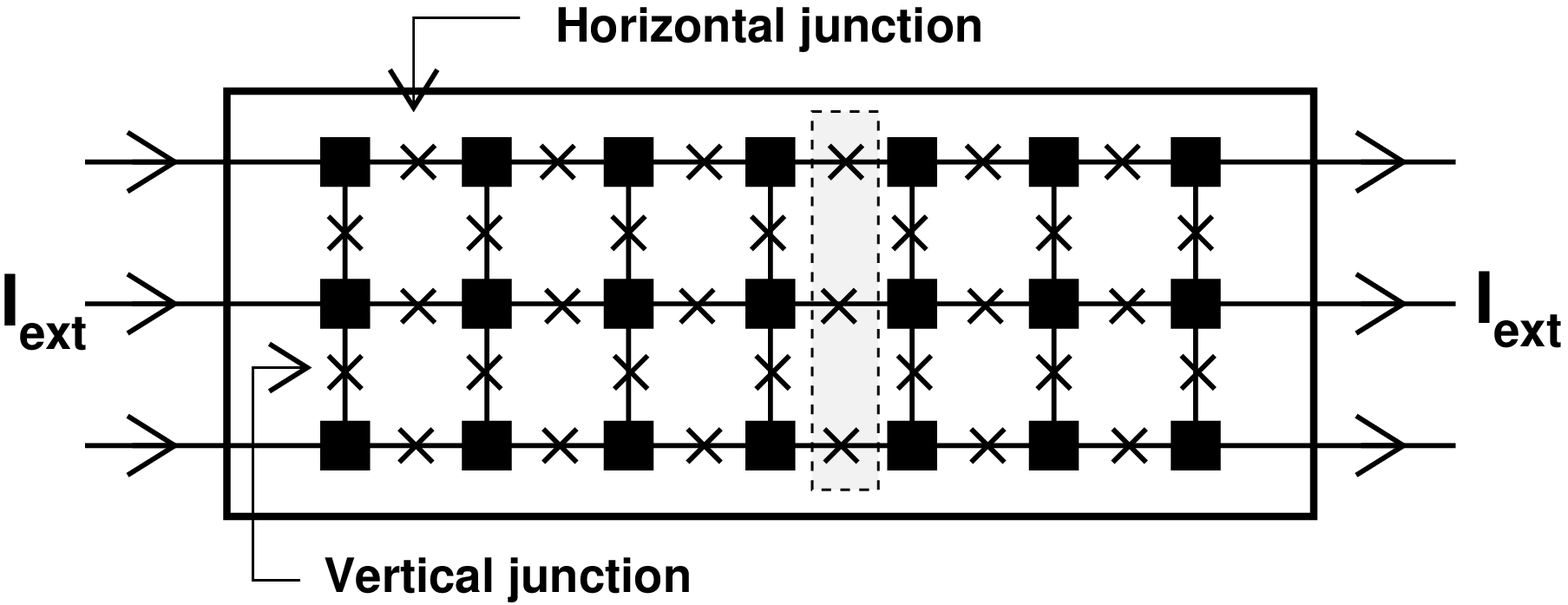}}
\caption{Sketch of the array geometry in our model.  There are
$(M\times N)$ superconducting islands (black squares), making
$\protect{[}(M-1)\times N+(N-1)\times M \protect{]}$ Josephson
junctions (crosses).  An external current $I^{ext}$ is injected into
each junction at one end of the array and extracted from each junction
at the other end.  The array is placed in an electromagnetic cavity
which supports a single resonant photon mode of frequency $\Omega$.
The dashes denote a ``row'' of junctions, which is perpendicular to
the current bias and is comprised of horizontal junctions.}
\label{fig:2Dgeometry}
\end{figure}

We consider a 2D array of $N \times M$ superconducting grains
placed in a resonant cavity, which we assume supports only a single
photon mode of frequency $\Omega$.  The array  thus has
$(N-1)(M-1)$ square plaquettes.  There are a total of $N_x \times N_y$
horizontal junctions, where $N_x = N-1$ and $N_y = M$.  A current $I$
is fed into each of the $M$ grains on the left edge of the array, and
extracted from each of the $M$ grains on the right edge.  Thus, the
current is injected in the $x$ direction, with no external current
injected in the $y$ direction.  A sketch of this geometry is shown in
Fig.\ \ref{fig:2Dgeometry}.  We also introduce the terminology that a
``row'' of junctions, in this configuration, refers to a group of
$N_y$ junctions, all with left-hand end having the same $x$
coordinate, and all being parallel to the bias current.  One such row
is indicated by the dashed lines in Fig.\ \ref{fig:2Dgeometry}.

We write the equations of motion for the grain variables (phases and
charges).  We express our Hamiltonian in the
form:\cite{almaas02_2,almaas03}
\begin{equation}
H = H_{photon} + H_J + H_C + H_{curr} + H_{diss}. \label{eq2d:ham}
\end{equation}
Here $H_{photon}$ is the energy of the cavity mode, expressed as
\begin{equation}
H_{photon} = \hbar\Omega\left(a^\dag a + \frac{1}{2}\right),
\end{equation}
where $a^\dag$ and $a$ as the usual photon creation and annihilation
operators.  $H_J$ is the Josephson coupling energy, which takes the
form
\begin{equation}
H_J = -\sum_{\langle ij \rangle} E^J_{ij}\cos(\gamma_{ij}),
\end{equation}
where $E^J_{ij}$ is the Josephson energy of the (ij)$^{th}$ junction,
and $\gamma_{ij}$ is the gauge-invariant phase difference across that
junction.  $E^J_{ij}$ is related to $I^c_{ij}$, the critical current
of the (ij)$^{th}$ junction, by $E^J_{ij} = \hbar I_{ij}^c /q$, where
$q = 2|e|$ is the Cooper pair charge.  $H_C$ is the capacitive energy
of the array, which we write as
\begin{equation}
H_C = \frac{1}{2}\sum_{ij}q^2(C^{-1})_{ij}n_in_j,
\end{equation}
where $C^{-1}$ is the inverse capacitance matrix, $n_i$ is the number
of Cooper pairs on the $i^{th}$ grain, and $q = 2e$ is the charge of a
Cooper pair (we take $e > 0$).  We write $\gamma_{ij}$ as
\begin{equation}
\gamma_{ij} = \phi_i - \phi_j - [(2\pi)/\Phi_0] \int_{ij} {\bf A}
	\cdot {\bf ds} \equiv~ \phi_i-\phi_j - A_{ij},
	\label{eq2d:gauge}
\end{equation} 
where $\phi_i$ is the gauge-dependent phase of the superconducting
order parameter on grain $i$, $\Phi_0 = hc/(2e)$ is the flux quantum,
and ${\bf A}$ is the vector potential, which in Gaussian units takes
the form\cite{slater,yariv}
\begin{equation}
{\bf A}({\bf x},t) = \sqrt{(h c^2) / (\Omega)} \left(a(t) +
       a^\dag(t)\right){\bf E}({\bf x}), \label{eq:A}
\end{equation}
where ${\bf E}({\bf x})$ is a vector proportional to the local
electric field of the mode, normalized such that $\int_Vd^3x|{\bf
E}({\bf x})|^2 = 1$. Again, $\Omega$ is the resonant frequency of the
cavity mode, and $V$ is the cavity volume.  The line integral is taken
across the (ij)$^{th}$ junction.  The phase factor $A_{ij}$ is then
\begin{equation}
A_{ij} = g_{ij} (a + a^\dag),
\end{equation}
where  
\begin{equation}
g_{ij} = \sqrt{\frac{\hbar c^2}{\Omega} \frac{(2\pi)^{3}} {\Phi_0^2}}
\int_{ij}{\bf E}\cdot{\bf ds}
\label{eq:gij}
\end{equation}
characterizes the effective coupling between the (ij)$^{th}$ junction
and the cavity.

One can also define a {\em frustration} $f_\mu$ for the $\mu^{th}$
plaquette by the relation
\begin{equation}
f_\mu = \frac{1}{2\pi}\sum_{plaquette}A_{ij},
\label{eq:fmu}
\end{equation}
where the sum runs over the bonds in the $\mu^{th}$ plaquette.  For
the present case,
\begin{equation}
f_\mu = f_\mu^{cavity} = 
\frac{1}{2\pi}(a + a^\dag)\sum_{plaquette}g_{ij}.
\label{eq:fmu1}
\end{equation}
If there were an applied magnetic field normal to the array in
addition to the cavity electric field, then the frustration would have
an additional contribution
\begin{equation}
f_\mu^{mag} = \Phi_\mu/\Phi_0, 
\label{eq:fmumag}
\end{equation}
where $\Phi_\mu$ is the magnetic flux through the $\mu^{th}$
plaquette.\cite{teitel}

Note that $g_{ij}$ is very sensitive to the experimental geometry.
For example, if the cavity has the form of a parallelepiped with edges
$L_x$, $L_y$, and $L_z$, where $L_z \geq L_x \geq L_y$, then lowest
mode in this cavity is a TE mode with frequency $\Omega = \pi
c\sqrt{1/L_z^2 + 1/L_y^2}$; the corresponding value of $g_{ij}$ is
given by\cite{almaas03}
\begin{equation}
g_{ij}^2 = e_{ij}^2\frac{32 e^2}{\hbar c}\frac{s^2}{L_y\sqrt{L_x^2 + 
L_z^2}}, \label{eq:value} 
\end{equation}
where $e_{ij}$ is the cosine of the angle between the ${\bf E}$ of the
resonant mode and the vector ${\bf ds}$).  In the geometry of Ref.\
\onlinecite{barbara}, $\Omega/(2\pi) \approx 100$ GHz, a cavity of
this shape gives rise to $g_{ij} \sim 0.001$,\cite{almaas03} though it
is obviously very sensitive to both array and cavity geometry.
 
The driving current and dissipation may be incorporated as
follows:\cite{almaas02_2,almaas03} The driving current is included via
a ``washboard potential,'' $H_{curr}$, of the form
\begin{equation}
H_{curr} = -\frac{\hbar I^{ext}}{q}\sum_{\langle ij\rangle \| \bf{ \hat{x}}}
	\gamma_{ij}, \label{eq2d:wash}
\end{equation}
where $I$ is the driving current injected in the $x$ direction into
{\em each} grain on the left edge and extracted from the right edge,
the sum running over bonds in the $x$ direction.  Dissipation is
included by coupling each $\gamma_{ij}$ to a separate collection of
harmonic oscillators with a suitable spectral
density.\cite{ambeg,caldeira81,leggett,chakra}
\begin{equation}
H_{diss} = \sum_{\langle ij \rangle} H_{ij}^{diss},
\end{equation}
where the sum runs over distinct bonds $\langle ij \rangle$, 
and
\begin{eqnarray}
H_{ij}^{diss} &=& \sum_{\alpha}\Bigg[f_{\alpha,ij}~ \gamma_{ij}~ u_{\alpha,ij} +
	\frac{(p_{\alpha,ij})^2}{2 m_{\alpha,ij}} \nn \\
	&+& \frac{1}{2}~ m_{\alpha,ij}~
	(\omega_{\alpha,ij})^2 ~(u_{\alpha,ij})^2 + \frac{(f_{\alpha,ij})^2}{
	2~ m_{\alpha,ij} ~(\omega_{\alpha,ij})^2}~ (\gamma_{ij})^2  \Bigg].
	\label{eq2d:diss}
\end{eqnarray}
The variables $u_{\alpha,ij}$ and $p_{\alpha,ij}$ of the $\alpha^{th}$
oscillator in the (ij)$^{th}$ junction, are canonically conjugate, and
$m_{\alpha,ij}$ and $\omega_{\alpha,ij}$ are the oscillator mass and
frequency.  If the spectral density, $J_{ij}(\omega) \propto
|\omega|$, the dissipation is ohmic.\cite{caldeira81,leggett} We write
such a linear spectral density as
\begin{equation}
J_{ij}(\omega) = \frac{\hbar}{2\pi}~ \alpha_{ij}~ |\omega|~ \Theta (\omega_c -
              \omega ),
\end{equation}
where $\omega_c$ is a high-frequency cutoff, $\Theta(\omega_c -
\omega)$ is a step function, and the dimensionless constant
$\alpha_{ij} = R_0/R_{ij}$, where $R_0 = h/(4e^2)$ and $R_{ij}$ is a
constant with dimensions of resistance, which turns out to be the
effective shunt resistance.

\section{Equations of motion}

The equations of motion are obtained by introducing operators $a = a_R
+ ia_I$ and $a^\dag = a_R - ia_I$.  These satisfy the commutation
relation $[a_R, a_I] = i/2$, which follows from $[a, a^\dag] = 1$.  In
terms of these variables, $H_{photon} = \hbar\Omega(a_R^2 + a_I^2)$ and
$\gamma_{ij} = \phi_i - \phi_j - 2 g_{ij} a_R$.

The time-dependence of the various operators appearing in the
Hamiltonian (\ref{eq2d:ham}) is now obtained from the Heisenberg
equations of motion.  These are readily derived from the commutation
relations for the various operators in the Hamiltonian
(\ref{eq2d:ham}).  Besides the relations already given, the only
non-zero commutators are $\protect{[}n_j, \phi_k\protect{]} = -i
\delta_{jk}$ and $\protect{[}p_{\alpha,ij}, u_{\beta,k\ell}\protect{]}
= -i\hbar~ \delta_{\alpha,\beta}~ \delta_{ij,k\ell}$, where the last
delta function vanishes unless $(ij)$ and $(k\ell)$ refer to the {\em
same} junction.

Using all these relations, we find, after a little algebra, the
following equations of motion for the operators $\phi_i$, $n_i$,
$a_R$, and $a_I$:
\begin{eqnarray}
\dot{\phi}_i & = &\frac{q^2}{\hbar}\sum_j(C^{-1})_{ij}n_j,\label{eq2d:eom1}\\
\dot{n}_i & = & -\frac{1}{\hbar} \sum_l E^J_{il} \sin(\phi_i-
             \phi_l- 2  g_{il} a_R) \nonumber \\ 
          &&+\frac{I_{i}^{ext}}{q} - \frac{1}{\hbar}\sum_{l}
             \sum_\alpha \left[u_{\alpha,il}f_{\alpha,il} +
	     \frac{(f_{\alpha,il})^2}{m_{\alpha,il}(\omega_{\alpha,il})^2}
	     (\phi_i-\phi_l -2  g_{il} a_R)\right],\label{eq2d:eom2}\\
\dot{a}_R &=& \Omega~ a_I,\label{eq2d:eom3}\\
\dot{a}_I &=&-\Omega~ a_R + \sum_{\langle ij \rangle}  g_{ij} 
	    \frac{E^J_{ij}}{\hbar}~ \sin(\phi_i-\phi_j - 2 g_{ij}a_R) - 
	    \frac{I^{ext}}{q} \sum_{\langle ij \rangle \|{\bf \hat{x}}}g_{ij} 
	    \nonumber\\
      && +\sum_{\langle ij \rangle} \frac{ g_{ij}}{\hbar}~ \sum_{\alpha} 
          \left(f_{\alpha,ij}~ u_{\alpha,ij} + \frac{(f_{\alpha,ij})^2}
	{m_{\alpha,ij} \omega_{\alpha,ij}^2}~ (\phi_i-\phi_j-2 g_{ij} 
        a_R) \right).\label{eq2d:eom4}
\end{eqnarray}
Here, the index $l$ ranges over the nearest-neighbor grains of $i$.
In writing these equations, we have assumed that the only external
currents $I_{i}^{ext}$ are those along the left and right edges of the
array, where they are $\pm I^{ext}$ [cf.\ Fig.\ \ref{fig:2Dgeometry}].
Eqs.\ (\ref{eq2d:eom1})-(\ref{eq2d:eom4}) are equations of motion for
the {\em operators} $a_R$, $a_I$, $n_j$, and $\phi_j$ (or
$\gamma_j$). In order to make these equations amenable to computation,
we will regard these operators as $c$-numbers. This should be
reasonable when there are many photons in the cavity.
\cite{almaas02_2,almaas03}

The equations of motion for the harmonic oscillator variables can also
be written out explicitly, but are of no interest; we instead
eliminate those variables and incorporate a dissipative term directly
into the equations of motion for the other variables.  Such a
replacement is possible provided that the spectral density of each
junction is linear in frequency, as above.  In that
case,\cite{ambeg,caldeira81,leggett,chakra,almaas02_2} the oscillator
variables can be integrated out.  The effect of carrying out this
procedure is that one should make the replacement $\sum_{\alpha}
\left( f_{\alpha,ij}~ u_{\alpha,ij} + \frac{(f_{\alpha,ij})^2}
{m_{\alpha,ij} \omega_{\alpha,ij}^2}~ \gamma_{ij} \right) \rightarrow
\frac{\hbar}{2\pi}\frac{R_0}{R_{ij}}~ \dot{\gamma}_{ij}$
wherever this sum appears in the equations of motion.  Making this
replacement in the equations of motion and simplifying, we obtain the
equations of motion for $n_j$ and $a_I$ with damping:
\begin{eqnarray}
\dot{n}_i &=& -\sum_j\frac{E^J_{ij}}{\hbar} \sin(\gamma_{ij}) + 
            \frac{I^{ext}_{i}}{q} -\sum_j\frac{1}{2\pi}\frac{R_0}{R_{ij}} 
            \dot{\gamma}_{ij}, \label{eq2d:eom2a} \\ 
\dot{a}_I &=& -\Omega~ a_R + \sum_{\langle ij \rangle} g_{ij} 
        \frac{E_{ij}}{\hbar}~ \sin(\gamma_{ij}) 
- \frac{I^{ext}}{q} \sum_{\langle ij \rangle \| {\bf \hat{x}}} 
g_{ij} + \sum_{\langle ij \rangle} g_{ij}~ \frac{R_0}{2\pi R_{ij}} \dot{\gamma}_{ij}.
\label{eq2d:eom4a}
\end{eqnarray}
Once, again, the index $j$ is summed only over the nearest-neighbor
grains of $i$.  Equations (\ref{eq2d:eom1}), (\ref{eq2d:eom3}),
(\ref{eq2d:eom2a}), and (\ref{eq2d:eom4a}) form a closed set of
equations which can be solved for the time-dependent functions
$\gamma_i$, $n_i$, $a_R$ and $a_I$, given the external current and the
other parameters.

To express these equations of motion in terms of suitable scaled
variables, we introduce a dimensionless time $\tau = t q R I^c /\hbar
= \omega_{\tau} t$, where $R$ and $I^c$ are suitable averages over
$R_{ij}$ and $I^c_{ij}$.  We also define the other scaled variables
$\tilde{R}_{ij} = \frac{R_{ij}}{R}$, $\tilde{\Omega} =
\frac{\Omega}{\omega_\tau}$, $\tilde{I} = \frac{I}{I^c}$, $\tilde{V}_i
= \frac{V_i}{R I^c}$, $\tilde{a}_{R,I}= \sqrt{2 \pi \frac{R}{R_0}}
a_{R,I}$, $\tilde{g}_{ij} = \sqrt{\frac{R_0}{2\pi R}} g_{ij}$,
$\tilde{C}_{ij} = \omega_{\tau} R C_{ij}$.  The last relation involves
the capacitance matrix $C_{ij}$.  We assume that this takes the form
\cite{fazio,kim} $C_{ij} ~=~\left(C_d + z_i~C_c \right) \delta_{ij} -
C_c \left(\delta_{i,j+{\bf \hat{x}}} + \delta_{i,j-{\bf \hat{x}}} +
\delta_{i,j+{\bf \hat{y}}} + \delta_{i,j-{\bf \hat{y}}}\right)$, i.\
e., that there is a non-vanishing capacitance only between neighboring
grains and between a grain and ground. Here $z_i (=4)$ is the number
of nearest neighbors of grain $i$, $C_d$ and $C_c$ are respectively
the diagonal (self) and nearest-neighbor capacitances, and ${\bf
\hat{x}}$ and ${\bf \hat{y}}$ are unit vectors in the $x$ and $y$
directions.  The corresponding Stewart-McCumber parameters are
$\beta_c = \omega_{\tau} R C_c$ and $\beta_d = \omega_\tau R C_d$.

In the above equations, we have introduced $V_i = q \sum_j
(C^{-1})_{ij} n_j$, which is the potential on site $i$.  The integral
of the electric field across junction $(ij)$ is $V_{ij} = V_i - V_j -
2 \tg_{ij} \Omega a_I$.

Carrying out these variable changes, and after some algebra, we find
the dimensionless equations of motion
\begin{equation}
\dot{\phi}_i      = \tilde{V}_i,
\label{eq:23}
\end{equation}
\begin{equation}
\dot{\tilde{V}}_i = \sum_j (\tilde{C}^{-1})_{ij} \Bigg[
	\tilde{I}_j^{ext} - \sum_l \Bigg( \tilde{I}_{j l}^c \sin (\phi_j - 
	\phi_l - 2 \tilde{g}_{j l} a_R ) \nn \\ + 
	\frac{1}{\tilde{R}_{j l}} (\tilde{V}_i
	- \tilde{V}_l - 2 \tilde{g}_{i l} \tilde{\Omega} \tilde{a}_I)\Bigg)
	 \Bigg],
\label{eq:24}
\end{equation}
\begin{equation} 
\dot{\tilde{a}}_R  = \tilde{\Omega} \tilde{a}_I, 
\label{eq:25}
\end{equation}
and
\begin{equation}
\dot{\tilde{a}}_I  =-\tilde{\Omega} \tilde{a}_R + \sum_{\langle ij 
	\rangle} \tilde{g}_{ij} \Bigg[ \tilde{I}_{i j}^c \sin (\phi_i\! -\! 
	\phi_j \!-\! 2 \tilde{g}_{i j} \tilde{a}_R) \nn \\ + 
	\frac{1}{\tilde{R}_{i j}} (\tilde{V}_i - \tilde{V}_j - 2
	\tilde{\Omega} \tilde{g}_{i j} \tilde{a}_I)\Bigg] - \tilde{I}^{ext} 
	\!\!\! \sum_{\langle ij \rangle \| {\bf \hat{x}}}\!\! 
\tilde{g}_{i j},
\label{eq:26}
\end{equation}
where the dot refers to a derivative with respect to $\tau$.  These
equations do not include the self-magnetic fields produced by the
currents.\cite{nakajima,majhofer,phillips,petraglia} However, they can
be generalized to include external currents with both $x$ and $y$
components, and non-square primitive cells.  Also, note that the
frustration parameter defined earlier is now time-dependent.

\section{Some Numerical Results}

We now describe some numerical results obtained by solving Eqs.\
(\ref{eq:23}) - (\ref{eq:26}) numerically,\cite{almaas03} using the
adaptive Bulrisch-Stoer method\cite{numrec} as further described in
Ref.\ \onlinecite{almaas02_2}.  For simplicity the coupling constants
$\tg_{ij}$ were assumed to have only two possible values, $\tg_x$ and
$\tg_y$, corresponding to junctions in the $x$ and $y$ direction
respectively. This assumption should be reasonable if there is little
disorder and the resonant mode has long wavelength compared to the
array dimensions.

\begin{figure}[pth]
\centerline{\includegraphics[height=13cm]{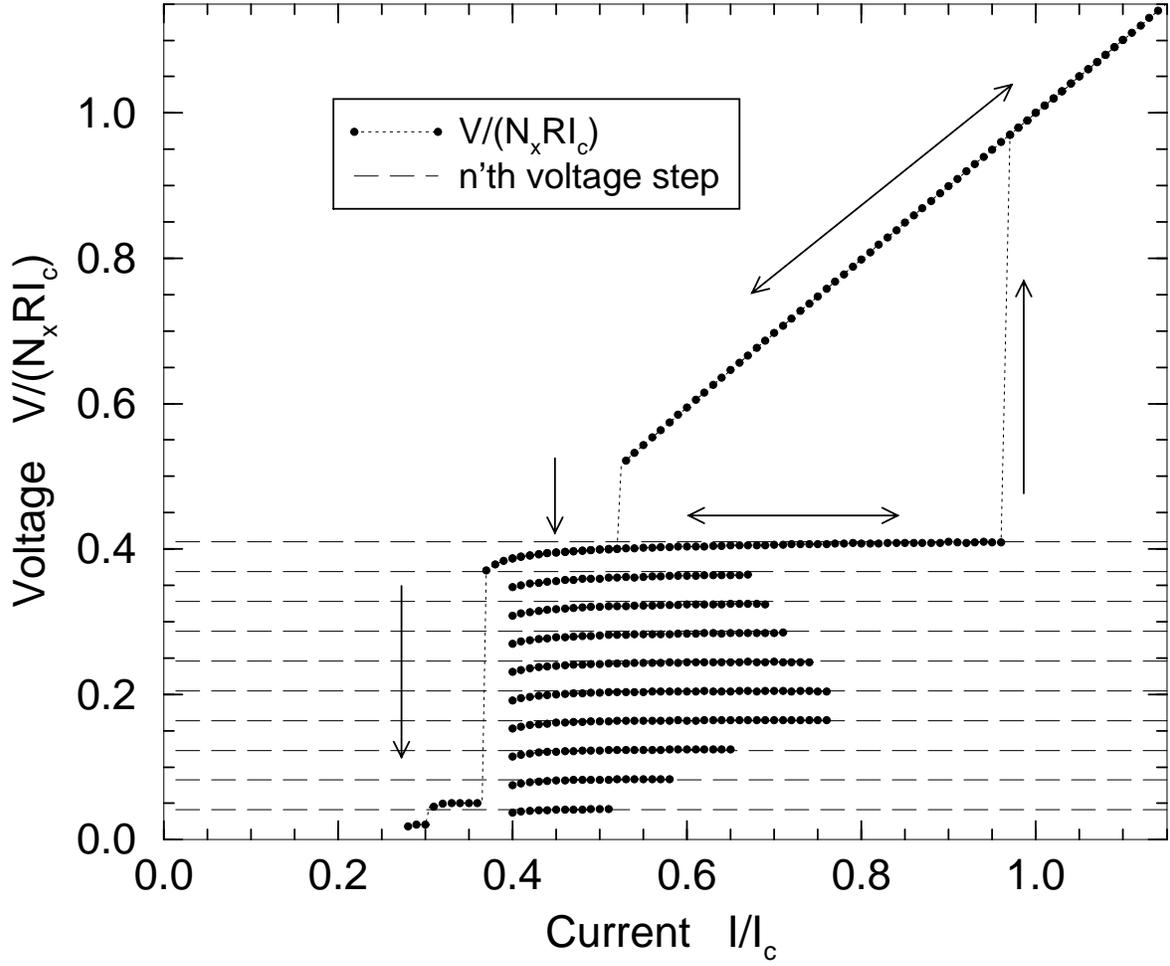}}
\caption{Calculated current-voltage characteristic for a $10\times4$
array with cavity frequency $\tilde{\Omega} = 0.41$, capacitance
parameters $\beta_c = 20$ and $\beta_d = 0.05$, disorder parameter
$\Delta = 0.05$ and junction-cavity coupling in the horizontal
direction $\tg_{x} = 0.012$. The horizontal dashed lines show expected
voltages for the various SIRS's.  These correspond to different
numbers of rows of horizontal junctions in the active state. Arrows
denote that the given IV was taken for increasing or decreasing
current.}
\label{fig:10x4_IV}
\end{figure}

The calculated IV characteristics for the case $\tg_x \neq 0$, $\tg_y
= 0$, with driving current parallel to the $x$ axis are shown in Fig.\
\ref{fig:10x4_IV}.  The array is taken to have $10\times 4$ grains,
with capacitances $\beta_c = 20$ and $\beta_d = 0.05$ (independent of
junction), $\tg_x = 0.012$, and $\tilde{\Omega} = 0.41$. The critical
current through the (ij)$^{th}$ junction is $\tilde{I}_{i j}^c = 1 +
\Delta_{ij}$ where the disorder $\Delta_{ij}$ is randomly selected
with uniform probability from $[-\Delta,\Delta]$, and $\Delta =
0.05$. The product $\tilde{I}_{i j}^c \tilde{R}_{ij}$ is assumed to be
the same for all junctions, in accordance with the Ambegaokar-Baratoff
expression.\cite{ambeg1} The calculated IV's are shown as a series of
points.  The arrow directions of the arrows indicate whether the
curves were obtained under increasing or decreasing current drive, or
both.  The horizontal dashed curves are voltages where {\em
self-induced resonant steps} (SIRS's) are expected, namely $\langle V
\rangle_{\tau}/(NRI_c) =\tilde{\Omega}$, where $\langle V
\rangle_{\tau}$ is the time-averaged voltage (dotted lines are guides
to the eye).  Each nearly horizontal series of points denotes a
calculated IV characteristic for a {\em different} number of active
rows $N_a$, and represents $N_a \times N_y$ (horizontal) junctions
sitting on the first integer ($n = 1$) SIRS.  The calculated voltages
agree well with expected values given by dashed horizontal lines.  The
long straight diagonal line segment represents the ohmic part of the
IV characteristic with all rows active.  (The corresponding segments
for other choices of $N_a < 10$ are not shown).  Besides the integer
SIRS's, there are a few fractional SIRS's, similar to what is seen for
Shapiro steps in conventional underdamped junctions.\cite{waldram}

\begin{figure}[pth]
\centerline{\includegraphics[height=13cm]{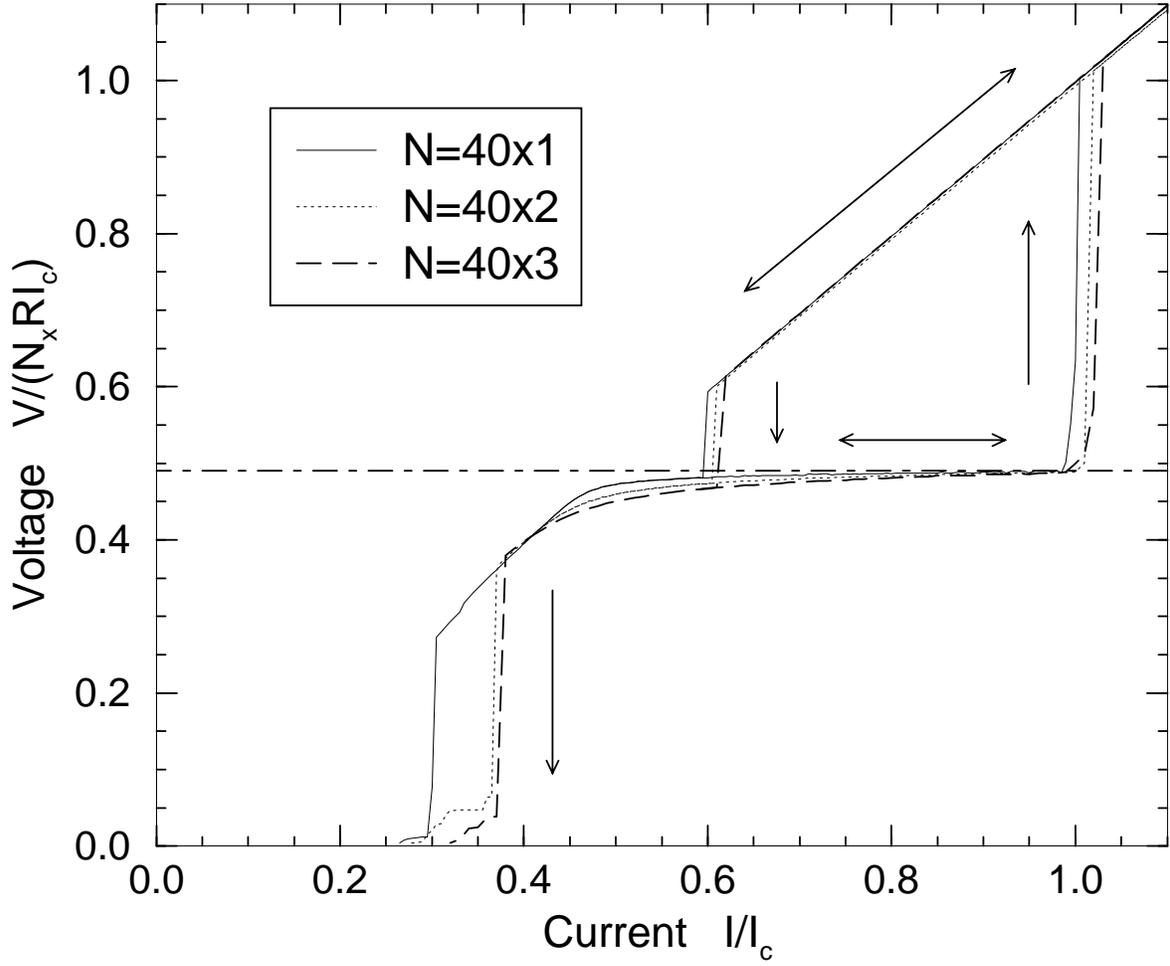}}
\caption{Calculated current-voltage characteristics for a $40\times 1$
(full line), a $40\times 2$ (dotted line) and a $40\times 3$
(long-dashed line) array, all with $\tilde{g}_x = 0.015$,
$\tilde{\Omega} = 0.49$, $\beta_c = 20$, $\beta_d = 0.05$ and $\Delta
= 0.05$.  The horizontal dot-dashed line shows the expected SIRS
position.  Note that as the array width increases, the smallest
$\tilde{I}$ at which all the active junctions phase-lock on the SIRS
also increases.  Hence, increasing the array width and increasing
$\tg_x$ have a similar effect.}
\label{fig:40x_IV}
\end{figure}

Fig.\ \ref{fig:40x_IV} shows the IV characteristics for three
different arrays, each with all rows in the active state: (i) a
$40\times 1$ (full curve), (ii) a $40\times 2$ (dotted curve) and
(iii) a $40\times 3$ (long-dashed curve).  Each array has the
parameters $\tilde{g}_x = 0.015$, $\tilde{\Omega} = 0.49$, $\beta_c =
20$, $\beta_d = 0.05$ and $\Delta = 0.05$. Once again, the arrows
denote the directions of current sweep.  The horizontal dot-dashed
curve shows the expected position of the SIRS corresponding to $N_a =
40$ [$V/(N_xRI_c) = \tilde{\Omega}$].  The curves show that all three
arrays have qualitatively similar behavior.  First, if the array is
started from a random initial phase configuration, such that
$\tilde{I} \equiv I/I_c > 1+\Delta$, and $\tilde{I}$ is decreased,
then all the rows lock on to the $N_a = 40$ SIRS.  Secondly, if
$\tilde{I}$ is further decreased, the $N_a = 40$ active state
eventually becomes unstable and all the junctions go into their
superconducting states.  Finally, if $\tilde{I}$ is {\em increased}
starting from a state in which the array is on the $N_a = 40$ SIRS,
the SIRS remains stable until $\tilde{I}$ reaches the critical current
for the various rows, and the IV curve becomes ohmic.

\begin{figure}[pth]
\centerline{\includegraphics[height=13cm]{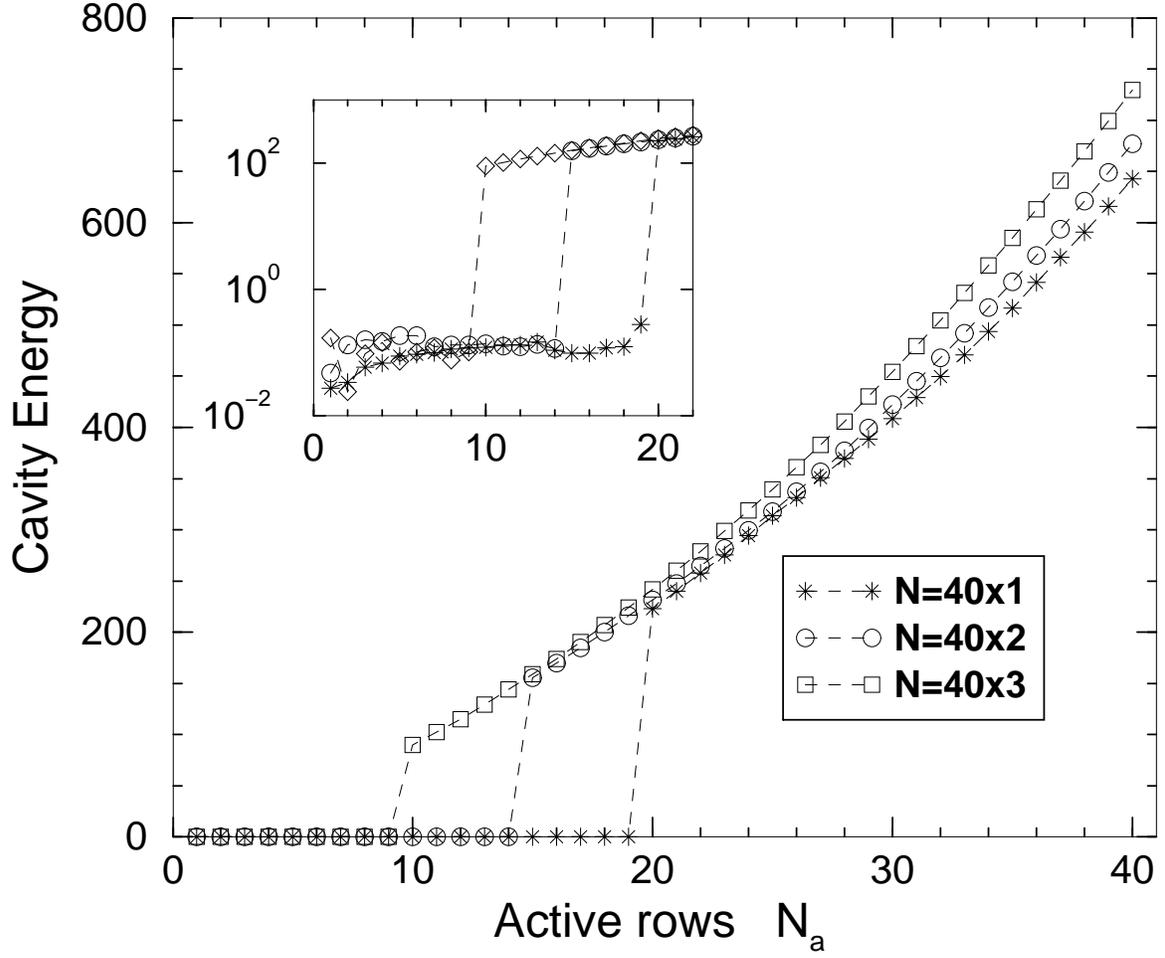}}
\caption{Time-averaged scaled energy $\tilde{E}$ in the resonant
cavity as function of active number of rows for a $40\times 1$
(asterisks), a $40\times 2$ (circles) and a $40\times 3$ (squares)
array with driving current $\tilde{I} = 0.58$. All the other
parameters are the same as those of Fig.\ \ref{fig:40x_IV}.  Inset: an
enlargement of the IV characteristics near the synchronization
threshold, on a logarithmic vertical scale.  The threshold number of
active junctions for synchronization decreases with increasing array
width.}
\label{fig:40x_thresh}
\end{figure}

In Fig.\ \ref{fig:40x_thresh}, we plot the time-averaged energy
$\tilde{E}(N_a) = \la \tilde{a}_R^2 + \tilde{a}_I^2 \ra_{\tau}$ in the
cavity for three different arrays: $40\times 1$ (stars), $40\times 2$
(circles), and $40\times 3$ (squares).  In all cases, $\tilde{I} =
0.58$, and the other parameters are the same as those of
Fig. \ref{fig:40x_IV}. Below a threshold value of $N_a$, (which we
denote $N_c$ and which depends on $N_y$), the active rows are in the
McCumber state (not on the SIRS's).  In this case, $\tilde{E}(N_a)$ is
small and shows no obvious functional dependence on $N_a$ (see inset).
By contrast, above threshold, $\tilde{E}(N_a)$ is much larger and
increases as $N_a^2$.  Fig.\ \ref{fig:40x_thresh} shows that, when
$N_y$ is increased at fixed $\tg_x$, $N_c$ decreases.  Precisely this
same trend is observed when we increase $\tg_x$ while holding $N_y$
fixed (and was observed in our previous 1D calculations with
increasing $\tg_x$).  Thus, the relevant parameter in understanding
the threshold behavior appears to be $N_y \tg_x$.

For the 2D arrays, one can introduce a define a {\em Kuramoto order
parameter} $\langle r_h\rangle_{\tau}$ for the horizontal bonds by
\begin{equation}
\langle r_h \rangle_{\tau} = \frac{1}{N_a N_y}~\langle | \sum_{\la ij \ra
     \parallel \hat{\bf x}} e^{i\gamma_{ij}}| \rangle_{\tau},
     \label{eq:2Dkuramoto}
\end{equation}
where $N_a$ is the number of active rows, $N_y$ is the number of
horizontal junctions in a single row and the sum runs over all the
active, horizontal junctions.  For the parameters shown in Fig.\
\ref{fig:40x_thresh}, it is found,\cite{almaas03} as in
1D,\cite{almaas02_2} that $\langle r_h \rangle_{\tau} \sim 1$ for $N_a
> N_c$ while $\langle r_h \rangle_{\tau} \ll 1$ for $N_a < N_c$.  This
behavior occurs because, for this choice of parameters, none of the
active junctions are on a SIRS when $N_a < N_c$.

For the case $\tg_x = 0$, $\tg_y \neq 0$, in our geometry (with square
primitive cells), we have not been able to find {\em any} value for
$\tg_y$ for which a SIRS develops.  This behavior is easily
understood.  In this geometry, with current applied in the $x$
direction, there is little power dissipated in the vertical junctions
and no resonance is induced in the cavity.

It is no surprise that the cavity interacts only very weakly with the
vertical junctions.  From previous studies of both underdamped and
overdamped disordered Josephson arrays in a rectangular geometry (see,
e. g., Refs.\ \onlinecite{wenbin92}, and \onlinecite{wenbin94}), it is
known that when current is applied in the $x$ direction, the $y$
junctions remain superconducting, with $\la V\ra_{\tau} \approx 0$,
while the $x$ junctions comprising an active row are almost perfectly
synchronized, with $\la r_x \ra \approx 1$.

If there were an an external magnetic field {\em perpendicular} to the
array, we believe that SIRS's would be generated for $\tg_y \neq 0$,
even if $\tg_x = 0$.  In this case, as mentioned earlier, there would
be a non-zero magnetic-field-induced frustration $f_\mu^{mag}$ [Eq.\
(\ref{eq:fmumag})]. As a result of this magnetic-field-induced
frustration, there would be nonzero voltages across the $y$ junctions,
as well as supercurrents in these junctions.

\begin{figure}[pth]
\centerline{\includegraphics[height=13cm]{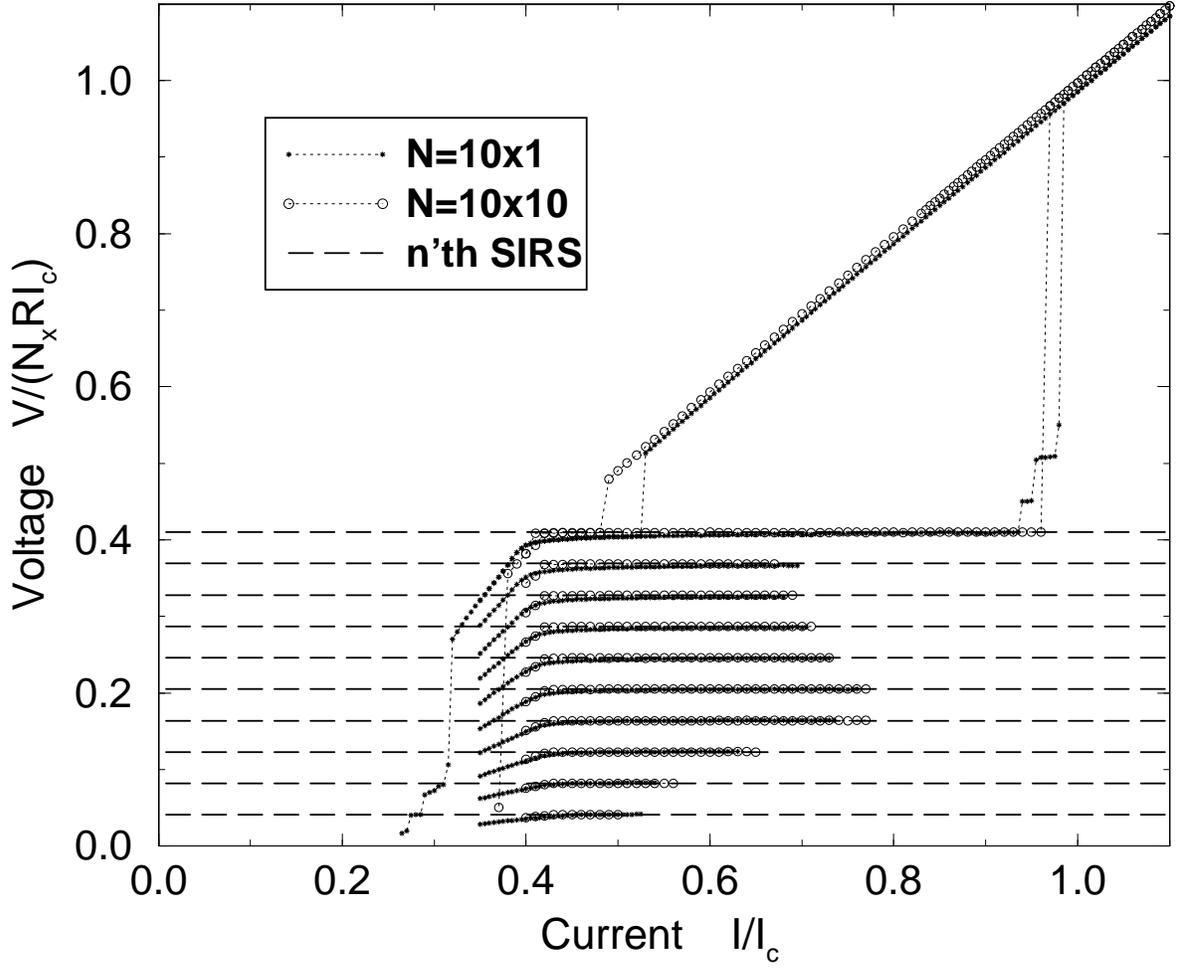}}
\caption{IV characteristics for a $10\times 1$ array ($\ast$) and a
$10\times 10$ array ($\circ$).  The $ 10 \times 1)$ array has
parameters $\tilde{g}_{x, 10\times 1} = 0.0259$, $\tilde{\Omega} =
0.41$, $\beta_c = 20$, $\beta_d = 0.05$ and $\Delta = 0.05$.  The
expected SIRS positions of the SIRS's are marked by horizontal dashed
lines.  The $10\times 10$ array has $\tg_{x, 10\times 10}$ = 0.00259,
and the other parameters are the same as for the $10\times 1 $ array.
The IV characteristics are shown for both increasing and decreasing
current drive.}
\label{fig:10x_IV}
\end{figure}

It is of interest to compare these 2D results explicitly with those of
1D arrays.  In Fig.\ \ref{fig:10x_IV}, the IV characteristics of a
$10\times 1$ array having coupling constant $\tg_{x;10\times 1} =
0.0259$ with those of a $10\times 10$ array with coupling constant
$\tg_{x;10\times 10} = 0.00259$.  The other parameters are the same
for the two arrays: $\tilde{\Omega} = 0.41$, $\beta_c = 20$, $\beta_d
= 0.05$ and $\Delta = 0.05$.  The expected positions of the SIRS's [at
$V/(NRI_c) = \tilde{\Omega}$] are indicated by dashed horizontal
lines.  Indeed, the two sets of IV characteristics are very similar.
This indicates that the crucial parameter is $N_y\tg_x$, not $N_y$ and
$\tg_x$ independently.  The slight extra flatness in the $10 \times
10$ IV's probably occurs because the individual junction couplings in
the $10 \times 1$ array are $10$ times larger than those in the $10
\times 10$ array.  Similarly, the differences in the ``retrapping
current'' in the two sets of curves (i. e. the current values below
which the McCumber curve becomes unstable), are due to the fact that,
for a given value of $\Delta$, the 2D arrays are effectively less
disordered than the 1D arrays, since the {\em average} critical
current for a single row has a smaller rms spread than the critical
current of a single junction in a 1D array.  (For other discussions of
the effects of disorder, see, e. g., Refs.\
\onlinecite{octavio,wiesenfeld1,wiesenfeld2}).

In both the $10\times 10$ and the $10\times 1$ array of Fig.\
\ref{fig:10x_IV}, the width of the SIRS's varies similarly (and
non-monotonically) with the number of active rows.  This behavior
distinguishes our predictions from those of some other
models,\cite{filatrella,filatrella2} where a monotonic dependence of
SIRS width on $N_a$ is found,\cite{barbara02} and the cavity is
modeled as an RLC oscillator connected in parallel to the entire
array.

\begin{figure}[pth]
\centerline{\includegraphics[height=13cm]{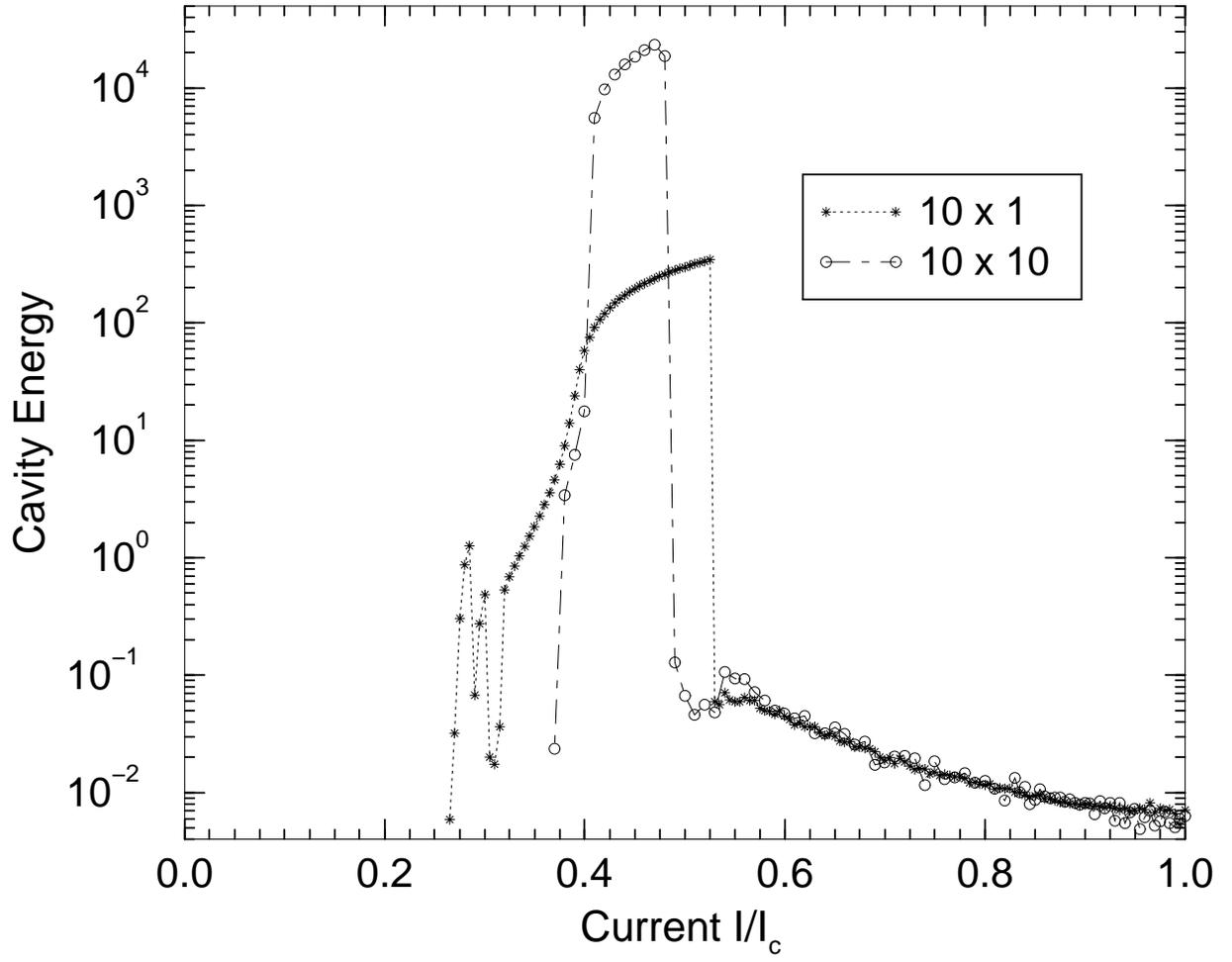}}
\caption{Time-averaged reduced cavity energy $\tilde{E}$, for a
$10\times 1$ array and a $10\times 10$ array for the same choice of
array parameters as in Fig.\ \ref{fig:10x_IV}.  The calculations are
carried out on the decreasing current branch with all rows active.
$\tg_x$ for the $10\times 10$ array is 10 times smaller than
that of the $10\times 1$ array.}
\label{fig:10x_cav}
\end{figure}

In Fig.\ \ref{fig:10x_cav}, the reduced time-averaged cavity energy
$\tilde{E} = \la a_R^2 + a_I^2 \ra_\tau$ is plotted vs.\ $\tilde{I} =
I/I_c$ for both arrays of Fig.\ \ref{fig:10x_IV}, under conditions
such that all rows are active.  This plot is obtained by following the
decreasing current branch.  When the $10\times 10$ array (with
$\tilde{g}_{(10\times 10)} = 0.1 ~\tilde{g}_{(10\times 1)}$) locks on
to the SIRS, $\tilde{E}$ jumps to a value which is approximately two
orders of magnitude {\em larger} than in the $10\times 1$ array, even
though $N_y\tg_x$ is the same for both arrays.  The reason is that
though the {\em width} of the steps is controlled primarily by
$N_y\tg_x$, the energy in the cavity is determined by the {\em square}
of the number of radiating junctions, which is is {\em 100 times
larger} for the 2D than the 1D array.

\section{Discussion and Summary}

The equations of motion described lead to a transition from
incoherence to coherence, as a function of the number of active rows
$N_a$.  There is also a striking effect of polarization: the
transition to coherence occurs only when the cavity mode is polarized
so that its electric field has a component parallel to the direction
of the current flow.

The numerical results closely resemble the experimental
behavior,\cite{barbara,vasilic02} showing the following experimental
features: (i) self-induced resonant steps (SIRS's) in the IV
characteristics; (ii) a transition from incoherence to coherence above
a threshold number of active junctions; and (iii) a total energy in
the cavity which varies quadratically with the number of active
junctions when those junctions are locked onto SIRS's.  But there may
be some differences, as discussed further in Ref.\
\onlinecite{almaas03}.

The 2D theory bears many similarities to the 1D case and elucidates
why the 1D model works so well.  These similarities occur because, in
a square array, only junctions which are {\em parallel} to the applied
current couple to the cavity.  Thus, as in 1D, the 2D model leads to
clearly defined SIRS's with voltages proportional to the cavity
resonant frequency.  However, there are some numerical results
specific to 2D.  For example, whenever one junction in a given row is
biased on a SIRS, {\em all} the junctions in that row phase-lock onto
that same SIRS.  In addition, when the array is biased on a SIRS,
$\tilde{E}(N_a)$ is much {\em larger} in 2D than in 1D, for the same
value of the coupling parameter $\tg_x N_y$.

A key difference between 1D and 2D is the effect of polarization: when
the cavity mode is polarized perpendicular to the applied current, it
does not affect the array IV characteristics.  We believe that, if the
array were frustrated, e.\ g. by an external magnetic field normal to
the plane of the array, there would be a coupling even when the cavity
mode is polarized perpendicular to the current.  It would be of
interest to carry out calculations for the model described here, to
confirm this effect.

We conclude with a brief discussion of ways in which our model might
be generalized to include some effects which are presently not taken
into account.  First, of course, one should include the fact that the
cavity, as well as the junctions, has a finite damping (finite $Q$).
Second, the real cavity fields are not spatially uniform within the
array, as assumed in the calculations presented here.  
We speculate that the effect of these two generalizations would be to
inhibit the array synchronization. Third, all real cavities have more
than one resonant mode; these other modes may be relevant in some
experimental circumstances.  Fourth, we have treated the operators as
$c$-numbers, i. e., have neglected quantum effects related to their
non-commutativity.  These quantum effects will certainly be relevant
in some circumstances.  Finally, the present treatment of damping
within individual Josephson junctions is also carried out in the
classical limit (resistively and capacitively shunted Josephson
junction); in the quantum regime where the individual junction
variables need to be treated quantum-mechanically, this treatment will
need to be modified.

Most of these generalizations can probably be carried out in a
straightforward manner.  Thus, we believe that the present model has
most of the essential physics underlying the SIRS's seen in
experiments.  Of the omitted effects, we believe that the quantum
effects, if included, are most likely to produce qualitative changes,
because they could lead to intriguing entanglement between quantum
states of the array and of the cavity.\cite{alsaidi}

\section{Acknowledgments}

This work has been supported by the National Science Foundation,
through grant DMR01-04987, and in part by the U.S.-Israel Binational
Science Foundation.  Some of the calculations were carried out using
the facilities of the Ohio Supercomputer Center, with the help of a
grant of time.  We are very grateful for valuable conversations with
Profs. T.~R. Lemberger and C.~J. Lobb.

\newpage

\end{document}